\documentclass[twocolumn,prd,aps,amsmath,floatfix]{revtex4} 

\usepackage{epsfig} 
\newcommand{\bea}{\begin{eqnarray}} 
\newcommand{\eea}{\end{eqnarray}} 
\newcommand{\Bx}{x_{bj}}  
\font\cmss=cmss12  
\def\1{\hbox{{1}\kern-.25em\hbox{l}}} 
\def\bfZ{\relax{\hbox{\cmss Z\kern-.4em Z}}} 
 
\begin{document} 
\def\lsim{\mathrel{\rlap{\lower4pt\hbox{\hskip1pt$\sim$}}
    \raise1pt\hbox{$<$}}}                
\def\gsim{\mathrel{\rlap{\lower4pt\hbox{\hskip1pt$\sim$}}
    \raise1pt\hbox{$>$}}}                

\title{Nuclear effects and their interplay in nuclear DVCS amplitudes} 
\author{A.~Freund}
\affiliation{Institut f{\"u}r Theoretische Physik, Universit{\"a}t Regensburg,  
D-93040 Regensburg, Germany} 
\author{M.~Strikman}
\affiliation{Department of Physics,
The Pennsylvania State University,\\
University Park, PA  16802, USA}
\medskip 

\begin{abstract} 
In this paper we analyze nuclear medium effects on DVCS amplitudes in
the $\Bx$ range of $10^{-1}-10^{-4}$ for a large range of $Q^2$ and four
different nuclei. We use our nucleon GPD model capable of describing
all currently available DVCS data on the proton and extend it to the
nuclear case using two competing parameterizations of nuclear
effects. The two parameterizations, though giving different absolute
numbers, yield the same type and magnitude of effects for the
imaginary and real part of the nuclear DVCS amplitude. The imaginary
part shows stronger nuclear shadowing effects compared to the
inclusive case i.e. $F^N_2$, whereas in the real part nuclear
shadowing at small $\Bx$ and anti-shadowing at large $\Bx$ combine
through evolution to yield an even greater suppression than in the
imaginary part up to large values of $\Bx$. This is the first time
that such a combination of nuclear effects has been observed in a
hadronic amplitude. The experimental implications will be discussed in
a subsequent publication.
\end{abstract} 
\maketitle 

\section{Introduction} 

Currently information about parton structure of nuclei is very
limited.  Most of the data were obtained at rather large $\Bx\geq
5\times10^{-2}$ where deviations of the nuclear structure from additivity
for $F_{2A}$ are of the order of a few percent, until one reaches the
region of the EMC effect $\Bx\geq0.4$ were the structure functions
themselves are already pretty small.  For gluons the situation is even
more uncertain - the theoretical analysis based on QCD momentum sum
rules and the expectation of a significant nuclear shadowing
\cite{FLS90,Eskola93} leads one to expect a gluon enhancement of the
order of $10-20\%$ for $\Bx\sim0.1$.  The analysis of the NMC data
supports such a conclusion \cite{Pirner,Eskola}.  However, the $Q^2$
range of the NMC data is rather limited and higher twist effects may
affect the analysis substantially.

The interpretation of the data at smaller $\Bx$ where one observes the
shadowing of $F_{2A}$, is even more difficult.  This is due to the
strong correlation of $\Bx$ and $Q^2$ in the relevant kinematic region
of $\Bx\leq10^{-2}$, $Q^2\leq2~\mbox{GeV}^2$. In this kinematic region
theoretical calculations of shadowing based on the Gribov theory
connecting diffraction in eN scattering and shadowing in eA scattering
describe the data well. However, in these analysis the vector meson
contribution which is a higher twist effect was found to be
substantial thus indicating the importance of higher twist shadowing
effects \cite{Piller,Thomas,Kaidalov}.  In principle, local duality
could result in the vector meson contribution being dual to the
continuum. However, the recent analysis in \cite{FGS03} based on the
theory of leading twist nuclear parton shadowing \cite{FS98,FGMS} has
explicitly demonstrated that $30-40\%$ of the shadowing for
$F_{2A}(\Bx\leq10^{-2}, Q^2\leq2~\mbox{GeV}^2)$ is due to higher twist
effects.

\begin{figure} 
\centering 
\mbox{\epsfig{file=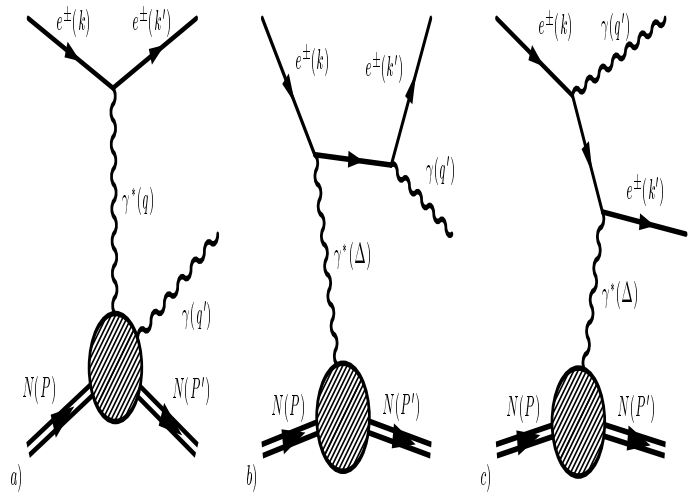,width=9cm,height=6.5cm}} 
\caption{(Color online) a) DVCS graph, b) BH with photon from final state lepton and c) with photon from initial state lepton.}
\label{dvcspic} 
\end{figure}

Clearly in this situation one needs new experimental tools to study
the nuclear partonic structure. Exclusive reactions appear to be a
very natural candidate. Among them deeply virtual Compton scattering
(DVCS) (see Fig.~\ref{dvcspic}) is the simplest reaction since the
final state real photon cannot be involved in the rescatterings, while
in the case of meson production a suppression of final state
interactions (color transparency) probably requires rather large
values of $Q^2$.

The most interesting production channel to investigate is the one
where the nucleus remains intact - coherent nuclear DVCS. The
generalized parton distributions (GPDs) which enter in this case carry
much more direct information about the nuclear structure than the
channels with a nuclear break-up.  Besides, the extraction of the
coherent signal is much easier experimentally both for fixed target
experiments with relatively modest resolution on the missing mass, and
for future collider experiments like the planned Electron Ion Collider
(EIC) where it will be possible to use a zero angle calorimeter to
select events were the nucleus remained intact.

This, however, comes at a price. The average values of $t$ in the
coherent scattering are small $\sim3/R_A^2$ and can hardly be measured.
Hence, one has to deal with cross sections integrated over $t$.  The
challenge in this case will be to observe the DVCS signal at small $t$
and small $\Bx$ where the QED Compton or Bethe-Heitler process is very
important.

Thus our task in studying nuclear DVCS is two fold. First, we want to
construct generalized nuclear parton distribution functions relevant
for the discussed process and then to calculate the DVCS amplitudes beyond
the impulse approximation which was considered in \cite{Guzey} for both
coherent and incoherent channels.  This will be the subject of the
first of two papers. In the second paper we will study various DVCS
observables to find out which of them are feasible for small $\Bx$
DVCS studies.

Our strategy will be to combine techniques which we developed in the
modeling of next-to-leading order (NLO) GPDs for nucleons in
\cite{afmmms} and which describe the current world DVCS data
\cite{HERAdvcs} with fairly high accuracy and the recent calculations
of nuclear PDFs at small $\Bx$ \cite{FS98,FGMS,FGS03}. These
calculations are based on the connection between diffractive parton
densities measured at HERA and shadowing for nuclear parton densities
\cite{FS98} and indicate that at small $\Bx$ shadowing should be large
in both the quark and gluon channels.

There are several aspects of nuclear DVCS at small $\Bx$ which appear
interesting. One goal would be to test our understanding of the
connection between PDFs and GPDs. Qualitatively, we expect that the
shadowing regime will start for the imaginary part of DVCS at larger
$\Bx$ since the effective $\Bx$ which enters in the DVCS amplitudes is
of the order of $\Bx/2$. The space-time picture of DVCS points in the
same direction since the coherence length in the transition of an
intermediate state of mass, $M$, into a real photon is larger than in
the case of DIS - the non-covariant propagator is $\sim 1/M^2$ instead of
$1/(M^2+Q^2)$.  Another aspect is the importance of the gluon GPD
feeding into the quark singlet GPD in the course of evolution
especially in the NLO approximation which was demonstrated in
\cite{afmmgpd}. Since the shadowing of gluons and quarks turns out to
be different within the model adopted in this paper, we expect the
A-dependence of the DVCS observables to be different from that of
$F_{2A}$.  For a comparison we will give also results based on the
model of \cite{Eskola} which is based on a fit to $\Bx\geq 10^{-2}$ data
assuming leading twist dominance and extrapolating to smaller $\Bx$
with the assumption that the gluon and quark parton densities are
screened in the same way at the initial scale of perturbative QCD
evolution.  Another unique aspect of DVCS is the ability to measure
the real part of the amplitude. In this case we expect significant
nuclear effects already at rather large $\Bx\simeq0.1$ since the real part
of the DVCS amplitude is sensitive to the rate of variation of parton
densities with $\Bx$, which, for the nuclear case, is already rather
different at $\Bx\sim5\times10^{-2}$.

Studies of nuclear DVCS will have important spin offs for studies of
color transparency in coherent high energy reactions since the same
GPDs enter in both DVCS and vector meson production. Hence comparing
results for these two processes will allow one to find the value of
$Q^2$ at which perturbative color transparency becomes
essentially model independent.

This paper is organized as follows: In section \ref{gpdmod} we
summarize the procedure for modeling nuclear GPDs based on the
information about nucleon GPDs.  We will also discuss the accuracy of
factorizing GPDs as a product of the GPD at $t=0$ and the nuclear form
factor.

In the next section we summarize the previous findings on NLO DVCS
amplitudes and give numerical results for a set of nuclei.  We will
demonstrate that nuclear effects are indeed quite strong and different
from the case of $F_{2A}$. We find that there is a significant
difference between nuclear effects between the two
cases. Interestingly enough we find that the nuclear effects for the
real part of the amplitude may extend all the way to the highest $\Bx$
where the coherent peak is still present, that is $\Bx\sim0.2$. This is
due to a curious interplay between nuclear shadowing and
anti-shadowing for $10^{-2}\leq \Bx \leq10^{-1}$.

In the second paper we will estimate various DVCS observables. The
main focus will be on the kinematic region of the EIC collider, though we
will also give numbers for fixed target kinematics. Overall we find that
DVCS signals are large enough for all standard observables to be
measured with high accuracy at the EIC.

\section{A nuclear GPD model and nuclear DVCS amplitudes}
\label{gpdmod}

In the following, we will use and review the model for twist-2 GPDs
first introduced in \cite{afmmms} using the off-diagonal
representation \cite{gb1} of GPDs.

Based on the aligned jet model (AJM) (see for example \cite{fs88}) the
key Ansatz of \cite{afmmms} in the DGLAP region is:
\begin{equation}
H^{S,NS,g} (X,\zeta) \equiv \frac{q^{S,NS,g}\left(\frac{X-\zeta/2}{1-\zeta/2}\right)}{1-\zeta/2}\, ,
\label{fwd}
\end{equation}
where $q^i$ refers to any forward distribution and $H^{S,NS}
(X,\zeta)= H^{q} (X,\zeta)\pm H^{\bar q} (X,\zeta)$. This Ansatz in
the DGLAP region corresponds to a double distribution model
\cite{rad1,rad2,rad3} with an extremal profile function allowing no
additional skewdness save for the kinematic one.

Note that we choose a GPD representation first introduced in
\cite{gb1}, which is maximally close to the inclusive case i.e
$X\in[0,1]$, $\zeta=\Bx$ with the partonic or DGLAP region in
$[\zeta,1]$ and the distributional amplitude or ERBL region in
$[0,\zeta]$. Note here that in the case of nuclei we speak of the
momentum fractions per nucleon, therefore, $\zeta\in[0,A]$.

The prescription in Eq.~(\ref{fwd}) does not dictate what to do in the
ERBL region, which does not have a forward analog. The GPDs have to be
continuous through the point $X=\zeta$ and should have the correct
symmetries around the midpoint of the ERBL region. They are also
required to satisfy the requirements of polynomiality:
\begin{align}  
M_N & = \int^1_{\zeta} \frac{dX \tilde X^{N-1}}{2-\zeta}\Big[H^q (X,\zeta)-(-1)^{N-1}H^{\bar q}(X,\zeta)\nonumber\\
& + \frac{(1+(-1)^{N})}{2}\tilde X H^g (X,\zeta)\Big] \nonumber \\
& +(-1)^{N}\int^{\zeta}_0\frac{dX \tilde X^{N-1}}{2-\zeta}\left[H^{\bar q}(X,\zeta) +\tilde X H^g (X,\zeta)\right]\nonumber\\
     &=  \sum^{N/2}_{k=0} \left(\frac{\zeta}{2-\zeta}\right)^{2k} C_{2k,N} \, , 
\label{polynomiality}  
\end{align}
with $\tilde X = \frac{X-\zeta/2}{1-\zeta/2}$.  The ERBL region is
therefore modeled with these natural features in mind. One demands
that the resultant GPDs reproduce the first moment $M_1= 3$ and
the second moment $M_2 = 1+C\zeta^2/(2-\zeta)^2$ \cite{footmom}.
$C$ was computed in the chiral-quark-soliton model \cite{vanderhagen1}
and found to be $-3.2$ and is related to the D-term \cite{poly} which
lives exclusively in the ERBL region.  This reasoning suggests the
following simple analytical form for the ERBL region ($X < \zeta$):
\begin{align}
&H^{g,NS}(X,\zeta) = H^{g,NS}(\zeta) \left[ 1+A^{g,NS}(\zeta) C^{g,NS} (X,\zeta) \right] \, , \nonumber \\
&H^{S}(X,\zeta) = H^{S}(\zeta)\left(\frac{X - \zeta/2}{\zeta/2}\right) \left[1 + A^{S}(\zeta) C^{S} (X,\zeta) \right] \, ,
\label{ajmerbl}
\end{align}
where the functions  
\begin{align}
&C^{g,NS} (X,\zeta) = \frac{3}{2}\frac{2-\zeta}{\zeta}\left(1 - \left( \frac{X-\zeta/2}{\zeta/2} \right)^2 \right)  \, , \nonumber \\
&C^{S} (X,\zeta)  = \frac{15}{2}\left(\frac{2-\zeta}{\zeta}\right)^2 \left(1 - \left(\frac{X-\zeta/2}{\zeta/2} \right)^2 \right) \, ,
\label{defCs}
\end{align} 
vanish at $X=\zeta$ to guarantee continuity of the GPDs.  The
$A^i(\zeta)$ are then calculated for each $\zeta$ by demanding that
the first two moments of the GPDs are explicitly satisfied. For the
second moment, what is done in practice is to set the D-term to zero
and demand that for each flavor the whole integral over the GPD is
equal to the whole integral over the forward input PDF without the
shift. For the final GPD, of course, the D-term is added to the
quark-singlet (there is no D-term in the non-singlet sector) using the
results from the chiral-quark-soliton model \cite{vanderhagen1}. The
gluonic D-term, about which nothing is known save its symmetry, is set
to zero for $Q_0$. Due to the gluon-quark mixing in the singlet
channel, there will be a gluonic D-term generated through evolution.

It would be straightforward to extend this algorithm to satisfy
polynomiality to arbitrary accuracy by writing the $A^i(\zeta)$
explicitly as a polynomial in $\zeta$ where the first few coefficients
are set by the first two moments and the other coefficients are then
either determined by the arbitrary functional form, as is done here,
or, perhaps theoretically more appealing, one chooses orthogonal
polynomials, such as Gegenbauer polynomials, for which one can set the
unknown higher moments equal zero. Phenomenologically speaking, the
difference between the two choices is negligible.

The above Ansatz also satisfies the required positivity conditions
\cite{rad2,posi1,posi2} and is in general extremely flexible both in
its implementation and adaption to either other forward PDFs or other
functional forms in the ERBL region. Therefore, it can be easily
incorporated into a fitting procedure making it phenomenologically
very useful. Since we will concern ourselves only with NLO, in what
follows we will use CTEQ6M \cite{cteq6} as the forward distribution.

In order to extend this parameterization for the proton to a nucleus,
we have to include the effects due to the nuclear medium. We do this
by multiplying the GPD in Eq.~(\ref{fwd}) with the ratio of nuclear
per nucleon to nucleon inclusive parton distribution function (PDF)
shifted by the same amount as the forward distribution in
Eq.~(\ref{fwd}):
\begin{align}
\frac{1}{A}H^{S,NS,g}_A (X,\zeta) = H^{S,NS,g}(X,\zeta)\cdot\frac{q^{S,NS,g}_A\left(\frac{X-\zeta/2}{1-\zeta/2}\right)}{A\cdot q^{S,NS,g}\left(\frac{X-\zeta/2}{1-\zeta/2}\right)}.
\label{nucgpd}
\end{align}
The ratio of nuclear to nucleon PDF is taken from \cite{FGS03},
\cite{Eskola}.  Note that the parametrization of Eq.\ref{nucgpd}
corresponds to the same value of $M_1$ as for the nucleon (which is
anyway necessary to satisfy the baryon charge conservation) and also
leads to an A-independent D-term. Note that it is difficult to
generate an A-dependent D-term via an attachment of the current to the
simplest t-channel exchanges. As an example consider the contribution
to the D-term due to two pions or a resonance in the t-channel. The
two-pion term where the pions are attached to two different
nucleons, is strongly suppressed for large A since it requires a spin
flip of both nucleons. Furthermore, it is quite difficult to attach a
resonance to two nucleons.  At the same time, one cannot exclude the
possibility of a coherent interaction of a current with several
nucleons. In fact, the real part of the amplitude appears to be more
sensitive to large distance effects which may enhance coherent
effects. Hence, one cannot exclude that the D-term actually does
depend on the atomic number\footnote{We thank M.Polyakov for a
  discussion of this point}. However, the quantities we consider are
not very sensitive to the D-term and so taking the D-term to be
independent of A seems to be a natural first approximation.

As far as the $t$-dependence is concerned, we postpone the discussion
to the next paper dealing with nuclear DVCS observables. Thus we will
consider the case $t=0$ for the time being. This is sufficient at
present, since larger values of $t>>t_{min}$ will not be
experimentally accessible in the coherent peak we are interested in.

After having evolved the GPDs using the same program successfully
employed in \cite{afmmgpd}, the real and imaginary part of the twist-2
DVCS amplitude in LO and NLO given below are calculated using the same
program as in \cite{afmmamp}:
\begin{align}  
&{\cal A}^{S}_{DVCS} (\zeta,\mu^2,Q^2) = \sum_a e^2_a \left(\frac{2 - \zeta}{\zeta} \right)\times\nonumber\\
&\Big[  
\int^1_0 dX~T^{S_a} \left(\frac{2X}{\zeta} - 1+i\epsilon, \frac{Q^2}{\mu^2} \right) ~{\cal F}^{S_a} (X,\zeta,\mu^2)  \nonumber\\  
&\Big. \mp \int^1_{\zeta} dX~T^{S_a} \left(1 - \frac{2X}{\zeta},\frac{Q^2}{\mu^2}\right)~{\cal F}^{S_a} (X,\zeta,\mu^2) \Big] ,
\end{align}
\begin{align}
&{\cal A}^{g}_{DVCS} (\zeta,\mu^2,Q^2) = \frac{1}{N_f}\left (\frac{2 - \zeta}{\zeta}\right )^2 \times\nonumber\\
&\Big[ 
\int^1_0 dX~T^{g} \left(\frac{2X}{\zeta} - 1+i\epsilon, \frac{Q^2}{\mu^2} \right) ~{\cal F}^{g} (X,\zeta,\mu^2)  \nonumber\\
&\pm \Big. \int^1_{\zeta} dX~T^{g} \left(1 - \frac{2X}{\zeta}, \frac{Q^2}{\mu^2}\right) ~{\cal F}^{g}(X,\zeta,\mu^2) \Big],  
\label{tdvcs}  
\end{align} 
where ${\cal F}$ stands for the nuclear GPD $H_A$ and ${\cal A}^i$ for the
respective Wilson coefficient functions.
 
The $+i\epsilon$ prescription is implemented using the Cauchy principal value
prescription (``P.V.'') through the following algorithm:
\begin{align} 
&\Big. P.V. \int^1_0 dX~T\left(\frac{2X}{\zeta} - 1\right) {\cal F}(X,\zeta,Q^2) =\nonumber\\
&\int^{\zeta}_0 dX~T\left(\frac{2X}{\zeta} - 1\right)\left({\cal F}(X,\zeta,Q^2)-{\cal F}(\zeta,\zeta,Q^2)\right) + \nonumber\\ 
& \int^1_{\zeta} dX~T\left(\frac{2X}{\zeta} - 1\right) \left({\cal F}(X,\zeta,Q^2) -{\cal F}(\zeta,\zeta,Q^2)\right)\nonumber\\
&+{\cal F}(\zeta,\zeta,Q^2)\int^1_0 dX~T \left(\frac{2X}{\zeta} - 1\right) \, . 
\label{subtraction} 
\end{align} 
The relevant LO and NLO coefficient functions are the same as for the
nucleon and can be found in
\cite{bemu3,afmmamp}.

\section{Results for nuclear DVCS amplitudes}
\label{nucdvcsamp}

In the following we give our results for the unpolarized nuclear DVCS
amplitudes in NLO for the following nuclei: $O-16$, $CA-40$, $Pd-110$
and $Pb-208$. We will use the FGS \cite{FGS03} and the Eskola et
al. \cite{Eskola} parameterizations of nuclear medium effects. The
initial scale for the perturbative evolution in both cases will be
$Q^2_0 = 2~\mbox{GeV}^2$. We choose such a low scale to be able to
make statements about HERMES kinematics ($0.05\leq\Bx\leq 0.3$ and
$1\leq Q^2\leq 9~\mbox{GeV}^2$) in our forthcoming paper on nuclear
DVCS observables. The FGS parameterization has an initial scale of
$Q^2_0 = 4~\mbox{GeV}^2$ and we, therefore, evolve it backward to our
$Q_0$. The parameterization of Eskola et al. has a $Q^2_0 =
2.25~\mbox{GeV}^2$ and is thus close enough to our starting
scale. Unfortunately, it is a LO parameterization which we will use in
NLO.

Our treatments of these two parameterizations can be rightfully called
into question. However, our aim at this point is not to make precision
statements about nuclear DVCS but rather look for the type of effects
which can be expected and how parameterization dependent they are. If
the differences in parameterizations will also materialize in DVCS
observables in a noticeable way, experiments we will have a good
chance of discriminating between different parameterization relying on
different assumptions of the underlying physics.

\begin{figure}
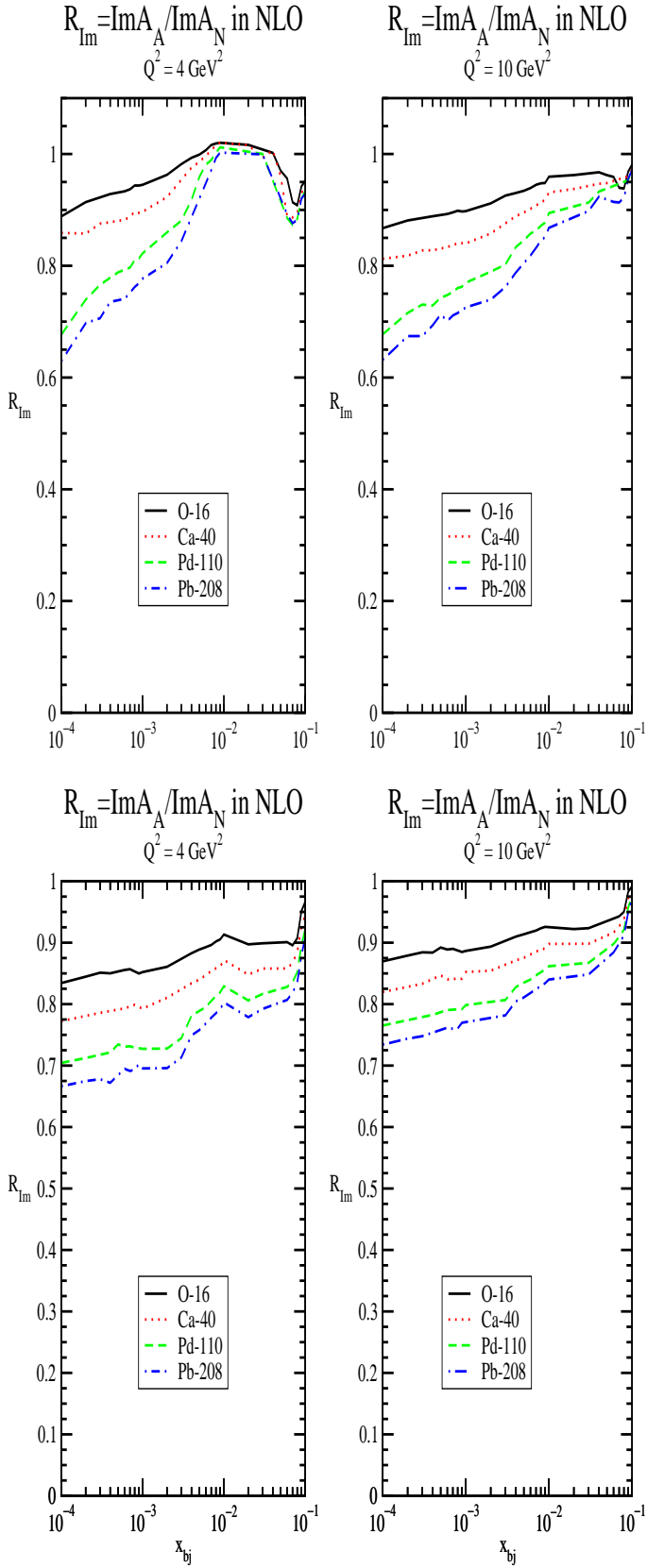
 
\centering 
\mbox{\epsfig{file=rationucleiampimnloq.eps,width=9cm,height=11cm}}
\mbox{\epsfig{file=rationucleiampimnloqesk.eps,width=9cm,height=11cm}}
\caption{(Color online) Ratio of nuclear per nucleon to nucleon DVCS amplitudes vs. $\Bx$ for two values of $Q^2$ for the FGS (upper plot) and Eskola (lower plot) parameterizations.}
\label{rat1}
\end{figure}

\begin{figure}
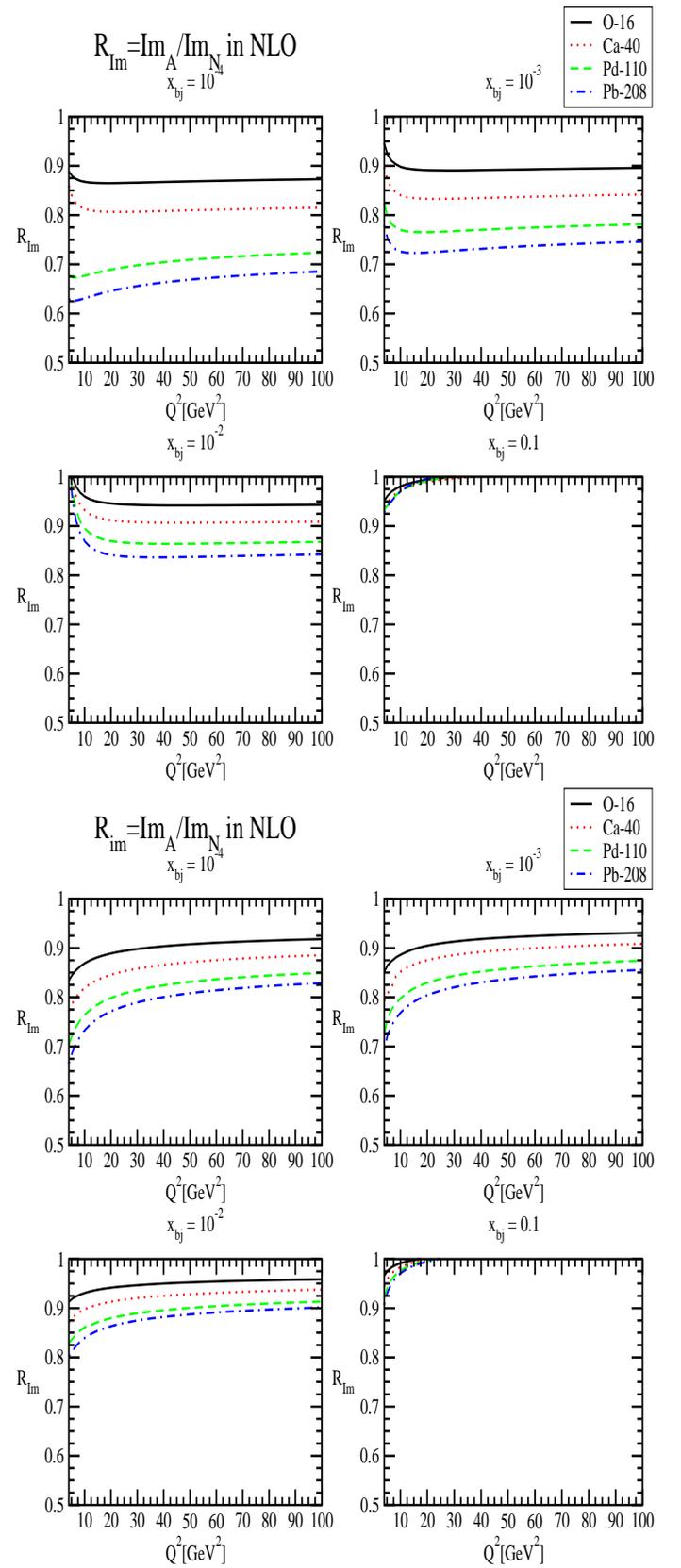
 
\centering 
\mbox{\epsfig{file=rationucleiampimnlox.eps,width=9cm,height=11cm}}
\mbox{\epsfig{file=rationucleiampimnloxesk.eps,width=9cm,height=11cm}} 
\caption{(Color online) Ratio of nuclear per nucleon to nucleon DVCS amplitudes vs. $Q^2$ for four values of $\Bx$ for the FGS (upper plot) and Eskola (lower plot) parameterizations.}
\label{rat2} 
\end{figure}

Let us first concentrate on the imaginary part of the unpolarized
nuclear DVCS amplitude which depends only on the DGLAP
region.  In Fig.~\ref{rat1}, we show the ratio of the imaginary part
of nuclear to nucleon DVCS amplitude in NLO for fixed $Q^2$ and
varying $\Bx$ and in Fig.~\ref{rat2} we show the same ratio for fixed
$\Bx$ and varying $Q^2$.

As is easily seen from Fig.~\ref{rat1}, the nuclear shadowing
corrections differ substantially between the two parameterizations at
low $Q^2$ and are very similar for larger $Q^2$. However, the nuclear
shadowing corrections are in both cases stronger than in the forward
case of $F^A_2/F_2$ (see for example \cite{FGS03}). In particular in
the region of small $\Bx\leq5\times10^{-3}$ one observes a significant slope
increase in $\Bx$ compared to the forward case. This behavior is
easily explained with first the shifted input toward lower values of
$X$. This leads to enhanced nuclear shadowing in the small $\Bx$
region since shadowing continues to increase toward smaller values of
$\Bx$. Secondly, since the value of the imaginary part of the
amplitude is mainly determined by the region $X\simeq O(\zeta)$ of large
light-like distances (see \cite{afgpd3d}) and since this region is
enhanced by perturbative evolution, one expects that until the nuclear
enhancement effects from large $X$ are degraded enough in momentum
through evolution, nuclear shadowing effects will increase with the
increase in $Q^2$. As can be seen from Fig.~\ref{rat2}, $Q^2$
evolution for fixed $\Bx$ will indeed initially increase nuclear
shadowing but the vanishing of nuclear shadowing with an increase in
$Q^2$ as seen in $F^A_2/F_2$ does not happen as rapidly as in DVCS.
This occurs precisely because the relevant region for the value of the
amplitude is the one of large light-like correlations $X\simeq O(\zeta)$ which
is enhanced through perturbative evolution. Large-light like
correlations in turn are more sensitive to nuclear medium effects and
thus nuclear shadowing will remain stronger for a wider range in
$Q^2$. We also observe a stronger $A$ dependence at small than at
large $\Bx$ which is again stronger than the forward case. Once more
the shift to relative smaller values of $\Bx$ is the reason. Nuclear
shadowing is stronger for $Pb-206$ than for $O-16$ and thus a relative
shift to smaller values of $\Bx$ will increase this difference
further. Evolution does not change this picture a lot since the
evolution effects are the same for all nuclei. As we just explained,
nuclear shadowing persists under evolution for a wider range in $Q^2$
than in the forward case where nuclear shadowing disappears more
rapidly and hence differences between nuclei as well.  When comparing
the two parameterizations one notices the peculiar behavior of the
ratio for the FGS parameterization at $Q^2=4~\mbox{GeV}^2$ in the
region of $5\times10^{-3}\leq\Bx\leq 10^{-1}$. This is due to the fact that in
the FGS parameterization the ratio of nuclear to nucleon gluon PDF is
above $1$ in the region $10^{-2}\leq\Bx\leq 10^{-1}$ and we are using
shifted distributions as our input for the GPDs. As it turns out the
imaginary part of gluon amplitude is negative whereas it is positive
for the quarks \cite{afmmamp} and the relative suppressions in the
quark and the gluon are sufficiently different enough in nuclei and
nucleons that in the sum this difference disappears! Note that this is
not the case for the Eskola parameterization. However, at
$Q^2=10~\mbox{GeV}^2$ this behavior has disappeared and the ratio is
the same for both distributions within $10-15\%$. The $A$ dependence of
the two parameterizations is quite different at lower $Q^2$, however,
quite similar at larger $Q^2$.

\begin{figure}
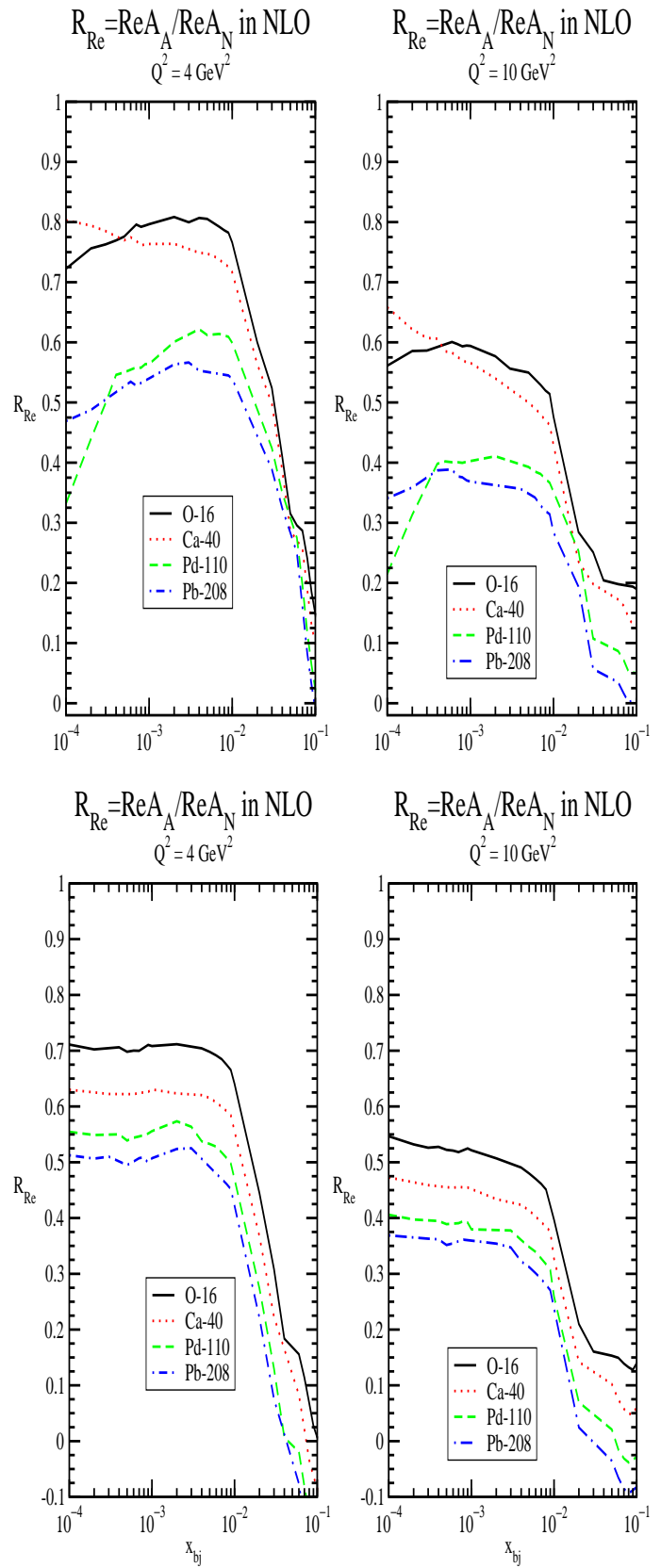
 
\centering 
\mbox{\epsfig{file=rationucleiamprenloq.eps,width=9cm,height=11cm}}
\mbox{\epsfig{file=rationucleiamprenloqesk.eps,width=9cm,height=11cm}} 
\caption{(Color online) Ratio of nuclear per nucleon to nucleon DVCS amplitudes vs. $\Bx$ for two values of $Q^2$ for the FGS (upper plot) and Eskola (lower plot) parameterizations.}
\label{rat3}
\end{figure}

\begin{figure}
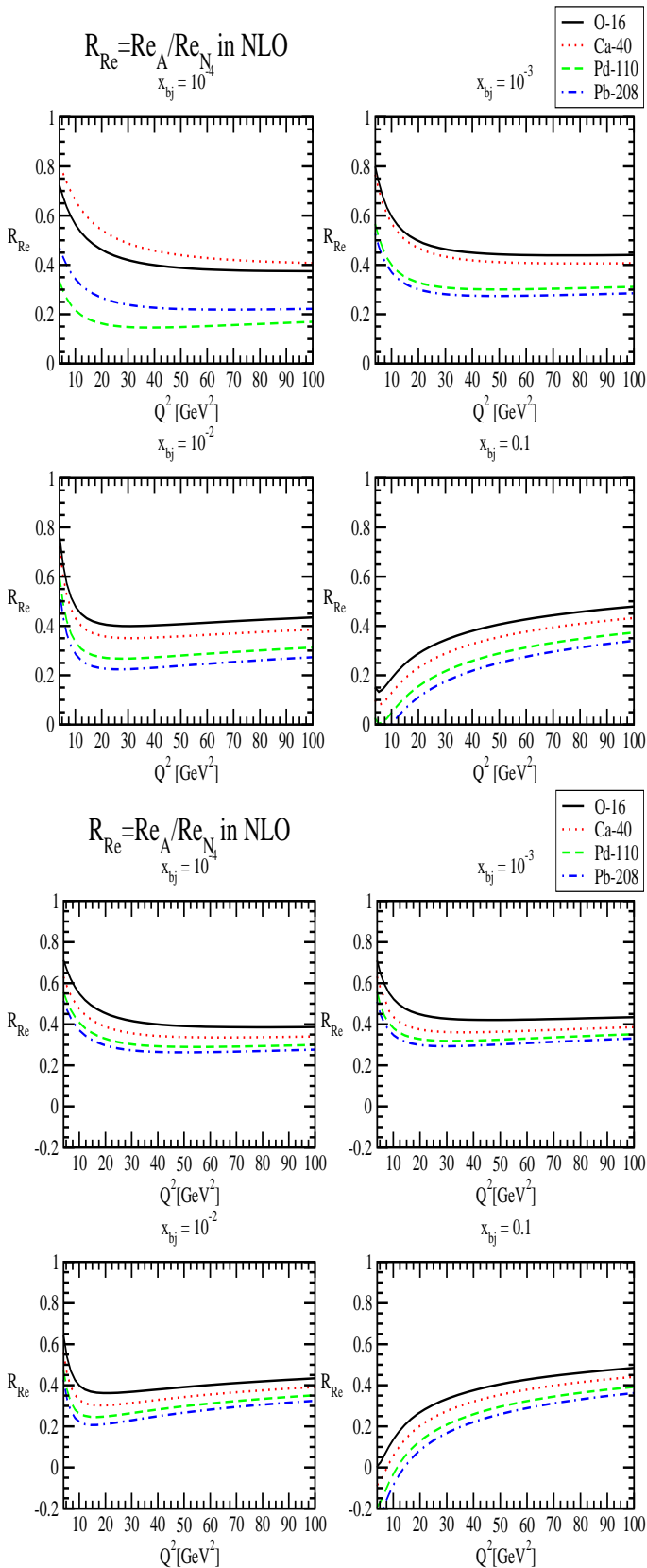
 
\centering 
\mbox{\epsfig{file=rationucleiamprenlox.eps,width=9cm,height=11cm}}
\mbox{\epsfig{file=rationucleiamprenloxesk.eps,width=9cm,height=11cm}} 
\caption{(Color online) Ratio of nuclear per nucleon to nucleon DVCS amplitudes vs. $Q^2$ for four values of $\Bx$ for the FGS (upper plot) and Eskola (lower plot) parameterizations.}
\label{rat4} 
\end{figure}

We now turn to the real part of the DVCS amplitude which depends both
on the ERBL and the DGLAP region. In Fig.~\ref{rat3}, we plot the
ratio of the real part of nuclear to nucleon DVCS amplitude in NLO for
fixed $Q^2$ and varying $\Bx$ and in Fig.~\ref{rat2} we show the same
ratio for fixed $\Bx$ and varying $Q^2$.

As can be readily seen in Fig.~\ref{rat3} the behavior of the real
part, for both parameterizations, is substantially different from the
one of the imaginary part for both parameterizations except for the
smallest values of $\Bx$. In particular the dip in the ratio in the
region $10^{-2}\leq\Bx\leq 10^{-1}$ is very surprising. The reason for this
suppression is quite deep. To put it in a nutshell, it is the
confluence of the nuclear medium effects: nuclear shadowing at small
$\Bx$ and anti-shadowing at large $\Bx$ which mix under evolution. How
is that possible?

The real part of the amplitude has two contributions: one originating
in the ERBL region and one in the DGLAP region with a relative minus
sign between them. If one were to decrease the contribution from the
ERBL region and increase the contribution from the DGLAP region, one
can obtain a significantly different value from the nucleon with just
a relative small change in both contributions due to the relative
minus sign. This is precisely what happens. To be definite let us take
$\Bx = 5\times10^{-2}$. At this value of $\Bx$ the value of the quark singlet is
virtually not shadowed with gluon already experiencing a stronger
suppression due to the relative shift to smaller values of $X$ where
nuclear shadowing is stronger. However, due to the nuclear enhancement
at larger $X$, the quark and gluon distributions in the ERBL are
smaller compared to the nucleon case in order to fulfill the first
and second moment of the nuclear GPD. At smaller $\Bx$ nuclear
shadowing leads to an additional suppression. All of this leads to a
smaller value of the contribution to the real part from the ERBL
region compared to the nucleon case. The contribution from the DGLAP
region increases since the value of the contribution due to the
nuclear enhancement increases. Therefore we have the required
configuration where the ERBL contribution decreases and the DGLAP one
increases. This does not drastically change under evolution though the
nuclear enhancement is pushed into the ERBL region the actual
evolution in the ERBL region is slower than in the DGLAP
region. Furthermore, the nuclear shadowing in the gluon leads to a
slower enhancement of the quark singlet compared to the nucleon case
leading basically to a preservation of the effect for a larger range
of $Q^2$ except for the large values of $\Bx$.

This can be observed in Fig.~\ref{rat4} where the ratio of the real
part is plotted vs. $Q^2$ for fixed $\Bx$ and the suppression effects
is seen to decrease slowly at small $\Bx$ and more rapidly at large
$\Bx$ though the absolute value of the effect remains large at large
$\Bx$ even for large $Q^2$. Again, the slow change in the ratio is
associated with the predominant enhancement of large light-like
distances.

This relative suppression of the ERBL region compared to the nucleon
case suggests the simple picture that whereas in the DGLAP region
partons could originate in principle from different nuclei, in the
ERBL this contribution is strongly suppressed. This once more has to
do with the fact that the ERBL region at small $\Bx$ is mainly
sensitive to large light-like distances and correlations over large
light-like distances are more susceptible to destructive nuclear
medium effects leading to a suppression of these parton correlations.

Since the suppression effect is very substantial, it should be clearly
measurable in a high precision $eA$ DVCS experiment at the EIC. We
will discuss this in detail in a forthcoming paper.

The above made observations would change somewhat if one were to
consider non-zero $t$ values. If we had performed a comparison at
$t=t_{min}=-\Bx^2m_N^2/(1-\Bx)$ rather than at $t=0$ the suppression
of the nuclear amplitude at $\Bx\ge 0.01$ would have been even stronger
(factor $2-5$ depending on the considered nucleus) since the nuclear
amplitude is suppressed by the nuclear form factor which can be
approximated as $F_A=\exp(R^2_At/6)$ for sufficiently small $t$, where
$R_A$ is the electromagnetic radius of the nucleus.

In practical applications considered in \cite{afmsdvcsnucob} the
effects of the nuclear form factor for asymmetries are approximately
cancelled in the ratio.

Furthermoew, for small $\Bx$ nuclear shadowing effects are likely to
lead to a small increase of the $t$-slope.  However, in the case of
cross sections integrated over $t$ (which is the only experimentally
measurable quantity) this effect leads to a reduction in the cross
section by no more than $10\%$ and it is even smaller for observables
sensitive to interference terms.

For $\Bx\gsim 0.1 (200/A^{1/3})$ deviations of the $t$-dependence from
that of the nuclear form factor are much more sensitive to the details
of the nuclear dynamics and require a more microscopic treatment than
the one presented here. It should be remembered that for $Pb-208$
$F_A(t_{min}(x=0.1))$ is close to the point where the form factor
changes sign and thus zero. Therefore, it will be extremely difficult to
extract the coherent cross section for such an $\Bx$ even using a veto for
production of forward neutrons.

When comparing the two parameterizations one finds a substantial
difference at low $Q^2$ and again similar values at large $Q^2$ except
for the smallest values of $\Bx$ where this difference persists over
the entire $Q^2$ interval. The most curious effect is seen at the
smallest values of $\Bx$ for the FGS parameterization where the
nuclear medium effect is less for $CA$ ($Pb$) compared to $O$
($Pd$). This is a peculiarity of the parameterization and once more the
extreme sensitivity of DVCS to the gluon distribution at small $\Bx$.
Note that the Eskola parameterization does not have this feature since
the gluon at small $\Bx$ is substantially different from the FGS
parameterization. Experiments will hopefully be able to distinguish
this effect in an observable sensitive to the real part like the beam
charge asymmetry.

\section{Conclusions}

In this paper we propose a model for nuclear GPDs based on the very
successful GPD parameterization used in \cite{afmmms} which describes
all available world data of DVCS on the nucleon. We calculate the
nuclear DVCS amplitudes using two different parameterizations for the
nuclear effects as proposed by \cite{FGS03} and \cite{Eskola} and plot
the ratio of nuclear to nucleon DVCS amplitude for both the real and
imaginary part. We find an enhancement of nuclear shadowing
in the imaginary part for small $\Bx$ compared to the inclusive case
which can be traced back to the high sensitivity of DVCS to large
light-like correlation distances which can be more easily influenced
by nuclear medium effects. In the real part we observe for the first
time an interplay between large and small $\Bx$ nuclear effects in
hadronic amplitudes for a large range in $Q^2$. This interplay is most
visible for $10^{-2}\leq\Bx\leq 10^{-1}$ where the effect is strongest and
should be observable in an $eA$ DVCS experiment as for example at the EIC.

\section*{Acknowledgment}

This work was supported by the DFG under the Emmi-Noether grant
FR-1524/1-3 and the DOE under grant number DE-FG02-93ER40771.


\begin{thebibliography}{99}
\bibitem{footmom} The first moment counts the number of quarks in the
  proton and the second moment is a generalization of the momentum sum
  rule where the D-term generates all the deviation from unity as $\zeta$
  varies.
\end{thebibliography}
\end{document}